\documentclass{article}
\usepackage{TGR}

\usepackage{amsmath,amsfonts,bm}

\def\eqref#1{equation~\ref{#1}}

\def\1{\bm{1}}

\DeclareMathAlphabet{\mathsfit}{\encodingdefault}{\sfdefault}{m}{sl}
\SetMathAlphabet{\mathsfit}{bold}{\encodingdefault}{\sfdefault}{bx}{n}

\usepackage{hyperref}
\usepackage{url}
\usepackage{graphicx}
\usepackage{multirow}
\usepackage{booktabs}
\usepackage{subcaption}
\usepackage{amsmath}
\usepackage{amssymb}
\usepackage{mathtools}
\usepackage{caption}
\usepackage{makecell}
\usepackage{wrapfig}

\newcommand\blfootnote[1]{%
  \begingroup
  \renewcommand\thefootnote{}\footnotetext{#1}%
  \endgroup
}

\newcommand{\first}[1]{\textbf{#1}}
\newcommand{\second}[1]{\underline{#1}}

\title{LoopGR: Parameter-Efficient Scaling for Generative Recommendation via Recurrent Depth}

\author{%
  Huizhe Zhang\textsuperscript{1,2,*}, 
  Yudong Li\textsuperscript{2,$\dagger$}, 
  Liang Chen\textsuperscript{1,$\dagger$}, 
  Chengxiang Zhuo\textsuperscript{2},
  Zang Li\textsuperscript{2}\\
  \normalfont 
  \textsuperscript{1}Sun Yat-sen University \quad
  \textsuperscript{2}Platform and Content Group, Tencent\\
  \normalfont zhanghzh33@mail2.sysu.edu.cn, chenliang6@mail.sysu.edu.cn \\
  \normalfont \{elsonli,felixzhuo,gavinzli\}@tencent.com
}

\begin{document}
\maketitle

\blfootnote{\textsuperscript{$*$}Work done during internship at Platform and Content Group, Tencent.}
\blfootnote{\textsuperscript{$\dagger$}Corresponding author.}

\begin{abstract}
Generative recommendation formulates item recommendation as autoregressive generation rather than a discriminative scoring problem. Although empirical studies exhibit favorable scaling behavior in recommendation, unleashing the benefits of scaling remains challenging under the strict model size and inference cost constraints of real-world systems. A central challenge is to combine expressive modeling of user-item interaction histories with flexible computational depth without proportionally increasing model size.
We introduce \textbf{LoopGR}, a parameter-efficient scaling framework for generative recommendation via recurrent depth. LoopGR employs a tailored hierarchical recurrent depth mechanism that alternates between a shared generative decoder and a lightweight context updater. For efficient inference, we further develop recurrent trajectory self-distillation that distills a low-budget student from a high-budget teacher by aligning both predictive distributions and evolution directions.
Extensive experiments show that LoopGR consistently outperforms strong generative recommendation baselines. Compared with a representative encoder–decoder baseline, LoopGR achieves superior performance while reducing model size by up to 52\%. These results establish recurrent depth as a new scaling dimension, offering a flexible and resource-efficient scaling path for generative recommendation.
\end{abstract}

\section{Introduction}

Generative recommendation, which reconceptualizes the traditional discriminative matching paradigm as a generative task, has been a promising paradigm for large-scale recommender systems~\citep{wang2023generative, zhou2025onerec, cheng2026tgr}. Inspired by tokenization in LLMs and computer vision, recent methods represent each item as a short sequence of discrete semantic IDs (SIDs) and formulate next-item recommendation as autoregressive generation of the target SIDs~\citep{rajput2023recommender, ju2026snapshot, he2026sidreasoner}. Empirical studies further suggest that increasing model capacity and training computation can improve recommendation performance, suggesting that generative recommenders also benefit from predictable scaling laws~\citep{guo2024scaling, liu2026understanding, hou2026survey}. However, unlike LLMs, recommender systems are deployed under much tighter computational and serving constraints. The key challenge is not simply to increase computation, but to effectively allocate a limited computational budget between interaction history modeling and target SID generation.

\begin{figure*}[t]
    \centering
    \includegraphics[width=\textwidth]{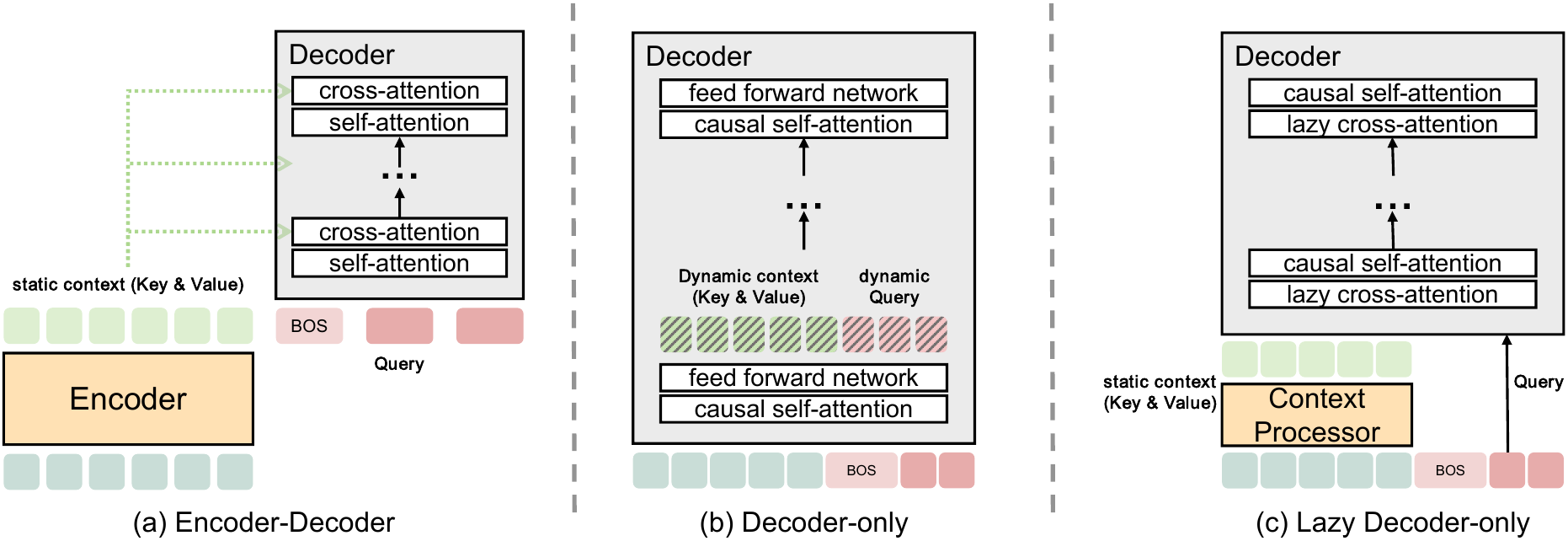}
    \caption{\textbf{Backbone comparison of different generative recommendation frameworks.} (a) Encoder-Decoder architectures allocate separate computation to history modeling and target generation. (b) Decoder-Only architectures unify this computation within a shared stack. (c) Lazy Decoder-Only architectures further reallocate the computational budget toward target generation.}
\label{fig:1}
\end{figure*}

Existing generative recommenders allocate computation differently between history modeling and target generation, as illustrated in Figure~\ref{fig:1}. This architectural evolution progressively shifts computation from dedicated history modeling toward shared and ultimately target-centric computation~\citep{rajput2023recommender, zhou2025onerecv2, yang2025cobra}. 
The above architectural designs reveal a common conflict between expressive context modeling and efficient target generation. Rich user-item context is typically obtained by deeply encoding historical interactions before generation, whereas efficient decoding often treats the resulting context as fixed conditioning. Meanwhile, computational depth remains tightly coupled with the number of network layers in existing architectures, such that deeper context refinement requires additional parameters. 
These observations motivate a central question: 

\begin{center}
\textit{How can we design a flexible generative recommendation backbone that dynamically organizes computation between context modeling and target SID generation?}
\end{center}

Looped Transformers, which recursively apply a set of shared layers across recurrent steps, offer a promising way to decouple computational depth from parameter size or memory consumption~\citep{yang2025nanbeige4, zhu2025scaling, geiping2026scaling}. Early studies combined recurrent depth with adaptive computation to iteratively refine token representations~\citep{dehghani2018universal, graves2016adaptive}, while recent work has revisited this paradigm to improve the reasoning and efficiency capabilities in modern language models~\citep{prairie2026parcae, wang2025hrm, saunshi2025reasoning}.
Although recurrent depth improves model efficiency via parameter sharing, it is hard to differentiate two distinct computations over long behavioral histories and target SIDs in generative recommenders. Moreover, additional recurrent steps still incur inference latency, making unrestricted looping impractical under strict serving budgets. This motivates a recommendation-specific recurrent design that separately organizes history refinement and target generation while supporting efficient low-budget inference.

In this paper, we propose \textbf{LoopGR}, a parameter-efficient scaling framework for generative recommendation via hierarchical recurrent depth. LoopGR is part of the TGR series of generative recommendation research at Tencent \cite{cheng2026tgr}. 
It comprises an encoder, a shared decoder and a lightweight shared context updater. After the encoder initializes the behavioral context, high-level decoder cycles and low-level updater cycles are recurrently interleaved. 
Specifically, the decoder predicts target SID representations from the current context while the updater feeds the emerging target state back to refine the context for subsequent decoding. This design allows user context to evolve during generation while scaling computational depth through shared layers. To support efficient inference under hard serving budgets, we propose a recurrent trajectory self-distillation to match predictive distributions and evolution directions from high-budget to low-budget recurrent trajectories. 
LoopGR reconciles dynamic context modeling with efficient target generation by decoupling computational depth from parameter count while retaining strong performance under limited recurrent budgets.
Our contributions are summarized as follows:
\begin{itemize}
    \item To the best of our knowledge, we are the first to introduce a recurrent-depth architecture for SID-based generative recommendation, which enables dynamic context refinement during target SID generation while decoupling computational depth from parameter count.
    \item  We propose a recurrent trajectory self-distillation method that aligns both predictive distributions and their evolution directions. It avoids the entanglement between exit selection and trajectory formation and effectively transfers a high-budget teacher into a low-budget student for efficient inference.
    \item Extensive experiments show that LoopGR outperforms strong baselines with fewer parameters. Beyond overall comparisons, we systematically evaluate recurrent depth, model size and computation scaling. Trajectory-level analyses are conducted to characterize the evolution of predictive representations across recurrent steps.
\end{itemize}

\section{Related Works}

\paragraph{Generative Recommendation.}
Generative recommendation has emerged as a new paradigm that directly generates target items from user-item interaction history~\citep{hou2026survey, li2025matching, wang2026ccformer}. With the rise of Large Language Models (LLMs), one line of studies adapts pretrained LLMs to recommendation by aligning their rich linguistic priors with recommendation-specific objectives~\citep{zheng2024adapting, onerecteam2026onereason, zhang2026latent}. For example, OneRec-Think~\citep{liu2025onerecthink} introduces itemic alignment, reasoning activation and reasoning enhancement to further equip recommenders with both explicit reasoning capabilities and personalized recommendation. 
A complementary line forgoes linguistic priors in LLMs and trains a recommendation-tailored architecture from scratch~\citep{rajput2023recommender, wang2025letter, wang2024eager}. OneRec-V2~\citep{zhou2025onerecv2} proposes a lazy decoder-only architecture to reallocate computation toward the target SID generation and avoid repeatedly applying expensive sequence encoding. Besides, several studies have further explored latent reasoning for generative recommendation by introducing intermediate latent states~\citep{yang2026reasoning, chen2026lasar, tang2026think}.
Despite these advances, most existing methods retain fixed-depth Transformer backbones, where increasing representational capacity typically entails introducing additional layers and parameters. Such parameter-centric scaling is difficult to reconcile with the stringent memory and latency requirements of recommendation systems.

\paragraph{Looped Transformers.}
Looped Transformers increase effective computational depth by repeatedly applying a set of shared layers to decouple the computational depth from model size~\citep{bai2019deep, lan2019albert}.
As an earlier study, Universal Transformers~\citep{dehghani2018universal} proposes a self-attentive recurrent sequence model with Adaptive Computation Time (ACT)~\citep{graves2016adaptive} to iteratively refine the symbol representations.
More recently, recurrent depth has been revisited in LLMs to improve parameter-efficient scaling and reasoning~\citep{yang2024looped, giannou2023programmable, fan2025looped}. As an exemplary implementation, Ouro~\citep{zhu2025scaling} repeatedly reuses a Transformer stack and systematically studies the scaling behavior of looped language models. Huginn~\citep{geiping2026scaling} adopts a different architecture in which the iterative body is tied between prelude and coda stacks. 
In recommender systems, recurrent depth remains largely underexplored. LoopCTR~\citep{tang2026loopctr} uses a parameter-shared prefix-attention module as the recurrent block for CTR prediction. However, it tailors recurrence layers to discriminative CTR prediction rather than autoregressive item generation. Deeper recurrence introduces additional sequential computation, making the balance between recurrent depth and inference efficiency an important direction for recommender systems.

\section{Preliminaries}

\paragraph{Semantic ID-based Generative Recommendation.}
Let $\mathcal{I}=\{1, \ldots, N\}$ denote the item set and $\mathcal{S}_u=(i_1, \ldots, i_n)$ the user-item interaction history of user $u$. The goal of next-item recommendation is to predict the subsequent item $i_{n+1}\in\mathcal{I}$. 
Instead of assigning each item an atomic identifier, semantic ID-based generative recommendation represents an item $i$ by a hierarchical code sequence $\mathbf{c}_i=(c_i^1,\ldots,c_i^H)$, where $H$ is the length of SIDs and $c_i^h$ denotes its code at hierarchy $h$. Following prior work~\citep{lee2022autoregressive}, each SID consists of semantic tokens generated by a Residual-Quantized Variational AutoEncoder (RQ-VAE) and an extra anti-collision token. Accordingly, next-item prediction over semantic IDs can be formulated as follows:
\begin{equation}
    p_{\theta}(i_{n+1}\mid\mathcal{S}_u)
    \equiv
    p_{\theta}(\mathbf{c}_{i_{n+1}}\mid\mathcal{S}_u)
    =
    \prod_{h=1}^{H}
    p_{\theta}\!\left(c_{i_{n+1}}^h
    \mid
    \mathbf{c}_{i_{n+1}}^{<h},\mathcal{S}_u\right),
\end{equation}
where $\theta$ denotes the parameters of the generative recommendation model.

\paragraph{Recurrent Depth.}
Looped architectures increase computational depth by repeatedly applying parameter-shared transformations. Given an initial hidden representation $\mathbf{H}^{(0)}$, the iterative computation can be written as
\begin{equation}
    \mathbf{H}^{(t+1)}=\mathcal{F}_{\phi}\!\left(\mathbf{H}^{(t)};\mathbf{E}\right), 
    \qquad 
    t=0,\ldots,T-1,
\end{equation}
where $\mathbf{E}$ denotes the conditioning context. $\mathcal{F}{\phi}$ denotes a recurrent transformation whose parameters $\phi$ are shared across iterations. $T$ denotes the number of recurrent steps, referred to as the loop budget. The intermediate states naturally form a trajectory in the latent space $\mathbf{\tau}=\left(\mathbf{H}^{(1)}, \mathbf{H}^{(2)}, \ldots, \mathbf{H}^{(T)}\right)$.

\section{Methodology}

\begin{figure*}[t]
    \centering
    \includegraphics[width=0.9\textwidth]{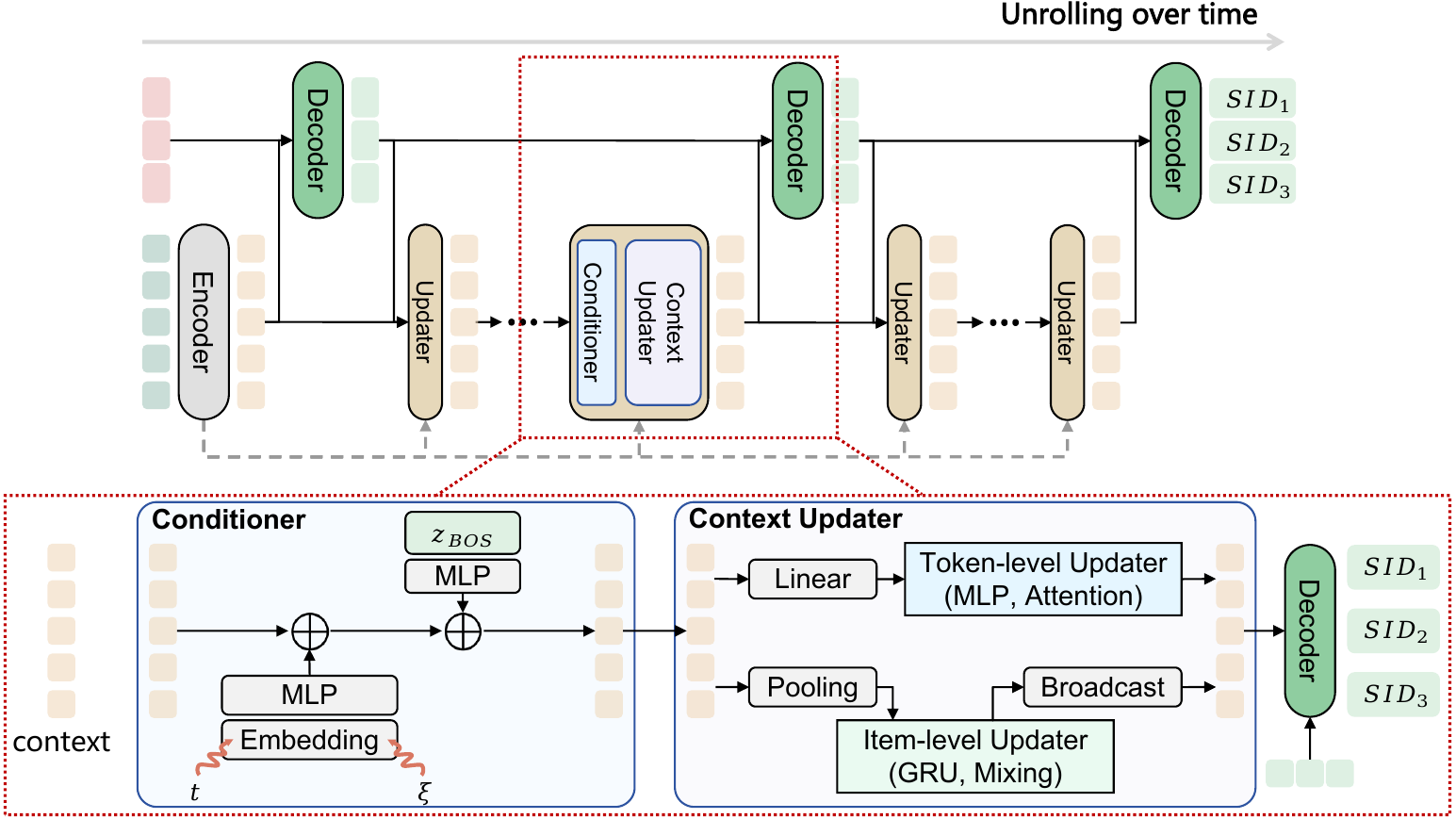}
    \caption{\textbf{Overview of LoopGR.} LoopGR alternates a shared decoder with a shared context updater to jointly refine target SID predictions and context representations over hierarchical recurrent depth.}
\label{fig:2}
\end{figure*}

In this section, we formally define LoopGR from two complementary perspectives, architectural design and training objective. As illustrated in Figure~\ref{fig:2}, LoopGR encodes the historical interactions into a contextual memory and then alternates parameter-shared decoding and lightweight context updates to progressively predict the target SID along recurrent depth. This recurrent computation naturally forms a multi-step prediction trajectory, which we further exploit through a tailored training objective that first optimizes a high-budget teacher and then distills its trajectory into a low-budget student. Consequently, LoopGR defines how a trajectory is formed, while the objective determines how this trajectory is exploited under different inference budgets for generative recommendation.

\subsection{LoopGR Architecture}

\paragraph{Overall Design.}
LoopGR organizes target generation as a hierarchical recurrent process consisting of an Encoder, a Context Updater and a Decoder. The hierarchical recurrent computation can be summarized as
\begin{equation}
\begin{aligned}
    \mathcal{F}
    =
    \operatorname{DEC}
    \circ
    \underbrace{
        \mathcal{B}\circ\cdots\circ\mathcal{B}
    }_{\text{high-level cycles}}
    \circ
    \operatorname{ENC},
    \qquad
    \mathcal{B}
    \coloneqq
    \underbrace{
        \operatorname{UPD}
        \circ\cdots\circ
        \operatorname{UPD}
    }_{\text{low-level cycles}}
    \circ
    \operatorname{DEC}.
\end{aligned}
\end{equation}
where $\circ$ denotes function composition. The high-level recurrence repeatedly revisits target prediction, while each high-level transition contains a nested low-level recurrence for progressively refining its conditioning context.

\paragraph{Encoder.}
We represent each historical item by its SIDs and obtain the corresponding token sequence $\mathcal{C}_u=(\mathbf{c}_{i_1},\ldots,\mathbf{c}_{i_n})$. After embedding these tokens into $\mathbf{X}_u\in\mathbb{R}^{L_u\times d}$, a Transformer encoder constructs the initial contextual memory:
\begin{equation}
    \mathbf{M}^{(0)}
    =
    \operatorname{ENC}\left(\mathbf{X}_u\right),
\end{equation}
where $L_u$ denotes the length of the historical SID token sequence and $d$ is the hidden dimension.

\paragraph{Context Updater.}
The context updater is introduced between decoding cycles to progressively refine historical representations via the previous prediction state. Specifically, at the $t_H$-th high-level cycle and the $t_L$-th low-level cycle, the updater constructs a cycle-dependent condition and extracts a prediction-aware query from the intermediate hidden states produced by the Decoder as follows: 
\begin{equation}
\begin{aligned}
    \mathbf{e}^{(t_H,t_L)}
    =
    \operatorname{MLP}
    \left(
        \operatorname{Emb}(t_H,t_L,\boldsymbol{\xi})
    \right),
    \qquad
    \mathbf{q}^{(t_H)}
    =
    \operatorname{MLP}
    \left(
        \mathbf{z}_{\mathrm{BOS}}^{(t_H)}
    \right),
\end{aligned}
\end{equation}
where $\mathbf{z}_{\mathrm{BOS}}^{(t_H)}$ denotes the hidden state of the BOS token after the $t_H$-th decoding cycle. As shown in Figure~\ref{fig:2}, the initial hidden states $\mathbf{Z}^{(0)}$ are obtained by passing the context embeddings $\mathbf{M}^{(0)}$ through the Decoder before the first context update. $\boldsymbol{\xi}$ denotes the hierarchy of SID.
These signals are injected into the context from the previous low-level step, and a lightweight sequence operator further refines the resulting conditioned representation:

\begin{align}
    \widetilde{\mathbf{M}}^{(t_H,t_L+1)}
    &=
    \operatorname{UPD}
    \left(
    \mathbf{M}^{(t_H,t_L)}
    ,
    \mathbf{e}^{(t_H,t_L)}
    ,
    \mathbf{q}^{(t_H)}
    \right),
    \\
    \mathbf{M}^{(t_H,t_L+1)}
    &=
    \operatorname{Norm}
    \left(
        \mathbf{M}^{(0)}
        +
        \widetilde{\mathbf{M}}^{(t_H,t_L+1)}
    \right),
\end{align}
 
In our implementation, the updater $\operatorname{UPD}(\cdot)$ shares parameters across all low-level cycles and is instantiated by default as a lightweight GRU-based operator. Notably, the updater is model-agnostic and we compare different instantiations in Section~\ref{sec:ablation_study}. 
Besides, the initial context embedding $\mathbf{M}^{(0)}$ provides a stable anchor for recurrent refinement to facilitate optimization over multiple low-level cycles. After $T_L$ low-level cycles, the context embedding is passed to the Decoder in the next high-level cycle and serves as the initial context for subsequent refinement:
\begin{equation}
\mathbf{M}^{(t_H+1,0)}
=
\mathbf{M}^{(t_H,T_L)}
=
\operatorname{Norm}
\left(
\mathbf{M}^{(0)}
+
\widetilde{\mathbf{M}}^{(t_H,T_L)}
\right),
\end{equation}

\paragraph{Decoder.}
The decoder generates recurrent target SID representations from the current historical context. At each autoregressive generation step, the same decoding procedure is repeated for $T_H$ high-level cycles before the resulting prediction is used for subsequent generation. Each decoding cycle is explicitly:
\begin{align}
    \widehat{\mathbf{Z}}^{(t_H)}
    &=
    \operatorname{Norm}
    \left(
        \mathbf{Z}^{(t_H)}
        +
        \mathbf{e}^{(t_H)}
    \right), \\
    \mathbf{Z}^{(t_H + 1)}
    &=
    \operatorname{DEC}
    \left(
        \widehat{\mathbf{Z}}^{(t_H)},
        \mathbf{M}^{(t_H+1, 0)}
    \right),
\end{align}
where $\mathbf{Z}$ denotes the decoder input formed by the BOS token and the target prefix. Similarly, $\mathbf{e}^{(t_H)}$ is constructed using the conditioning mechanism introduced above with $t_H$ as the only cycle index. $\mathbf{M}^{(t_H, 0)}$ denotes the contextual memory used at the $t_H$-th high-level cycle.

\subsection{Recurrent Trajectory Self-Distillation}

Increasing recurrent depth allows LoopGR to refine its predictions without increasing the parameter count, but deeper unrolling inevitably introduces additional sequential inference cost. Therefore, we propose recurrent trajectory self-distillation to transfer the refinement dynamics of a high-budget LoopGR to a low-budget counterpart.
Let $T^{tr}_H$ and $T^{stu}_H$ denote the numbers of high-level cycles used by the teacher and student, where $T^{stu}_H < T^{tr}_H$. We first train a high-budget teacher via the last step supervision.
We initialize the student from the trained teacher, $\theta_S^{(0)}\leftarrow\theta_T^{\star}$, and reduce the recurrent budget to $T^{stu}_H$ cycles. The above training objective can be formulated as follows:
\begin{equation}
\begin{aligned}
    \underbrace{
    \theta_T^{\star}
    =
    \arg\min_{\theta_T}\mathcal{L}_{\mathrm{task}}^{T}
    }_{\text{Stage I}},
    \qquad
    \underbrace{
    \theta_S^{\star}
    =
    \arg\min_{\theta_S}
    \left(
    \mathcal{L}_{\mathrm{task}}^{S}
    +\alpha\mathcal{L}_{\mathrm{Traj}}
    +\beta\mathcal{L}_{\mathrm{Der}}
    \right)
    }_{\text{Stage II}},
\end{aligned}
\end{equation}
where $\mathcal{L}_{\mathrm{task}}$ denotes the next-token prediction loss, and $\alpha$ and $\beta$ are loss weights. At each high-level cycle, the decoder hidden states are projected to token logits and converted into predictive distributions $\mathbf{p}$ via softmax. We align the student's predictive distributions with those at the corresponding cycles along the high-budget teacher trajectory as follows: 
\begin{equation}
     \mathcal{L}_{\mathrm{Traj}}
    =
    \frac{1}{T_H^{\mathrm{stu}}H}
    \sum_{s=1}^{T_H^{\mathrm{stu}}}
    \sum_{h=1}^{H}
    \mathcal{D}_{\mathrm{KL}}
    \left(
        \mathbf{p}_{\mathrm{tr}}^{(a(s),h)}
        \,\Vert\,
        \mathbf{p}_{\mathrm{stu}}^{(s,h)}
    \right).
\end{equation}
where $a(s)=\left\lceil\frac{sT^{tr}_H}{T^{stu}_H}\right\rceil$. It maps each student cycle to the corresponding state along the teacher trajectory. Following prior trajectory distillation~\citep{gong2025beyond}, we further minimize the directional discrepancy based on the first-order change:
\begin{equation}
    \mathcal{L}_{\mathrm{Der}}
    =
    \frac{1}{(T_H^{\mathrm{stu}}-1)H}
    \sum_{s=2}^{T_H^{\mathrm{stu}}}
    \sum_{h=1}^{H}
    \mathcal{D}_{\mathrm{cos}}
    \left(
        \Delta\mathbf{z}_{\mathrm{tr}}^{(a(s),h)},
        \Delta\mathbf{z}_{\mathrm{stu}}^{(s,h)}
    \right),
\end{equation}
 The finite differences in log-probability space are defined as $\Delta\mathbf{z}_{S}^{(s,h)} = \log\mathbf{p}_{S}^{(s,h)} - \log\mathbf{p}_{S}^{(s-1,h)}$ and $\Delta\mathbf{z}_{T}^{(a(s),h)} = \log\mathbf{p}_{T}^{(a(s),h)} - \log\mathbf{p}_{T}^{(a(s-1),h)}$.
Consequently, the student learns to approximate the high-budget refinement process with substantially fewer loop steps.

\section{Experiments}

\subsection{Experimental Setup}

\paragraph{Datasets.}
We conduct experiments on three widely used categories from the Amazon Reviews 2014 dataset~\citep{mcauley2015image}: \textit{Sports and Outdoors} (Sports), \textit{Beauty} (Beauty) and \textit{CDs and Vinyl} (CDs). We treat user reviews as implicit interactions and sort them chronologically to construct user behavior sequences. For the training set, we construct next-item prediction instances using sliding windows over each user's interaction history~\citep{rajput2023recommender}. We also evaluate LoopGR on an industrial dataset collected from Tencent's production recommendation system and compare it with two generative recommendation baselines deployed in Tencent's real-world services. More details about the datasets are provided in Appendix~\ref{apdx:1}.

\paragraph{Baselines.}
We compare LoopGR with representative recommendation methods from two categories. (i) Conventional recommendation methods include \textsc{GRU4Rec}~\citep{hidasi2015session}, \textsc{SASRec}~\citep{kang2018self}, \textsc{BERT4Rec}~\citep{sun2019bert4rec},  \textsc{FDSA}~\citep{zhang2019feature}, \textsc{S$^3$-Rec}~\citep{zhou2020s3} and \textsc{VQ-Rec}~\citep{hou2023learning}.
(ii) Generative recommendation methods include \textsc{TIGER}~\citep{rajput2023recommender},  \textsc{HSTU}~\citep{zhai2024actions}, \textsc{ActionPiece}~\citep{hou2025actionpiece} and \textsc{RPG}~\citep{hou2025generating}. 
We report the best-performing full-budget \textsc{LoopGR}, together with \textsc{LoopGR-RTSD}, its corresponding student operating at half of the recurrent budget.
For the industrial dataset, we compare LoopGR with TIGER, \textsc{BARGE}~\citep{zeng2026barge} and a plain Lazy Decoder-Only model derived from OneRec V2~\citep{zhou2025onerecv2}. Notably, the encoder or context processor of each baseline is adapted to accommodate the industrial input format in our implementation.

\paragraph{Implementation Details.}
All experiments are conducted on four NVIDIA H20 GPUs. For SID construction, we generate the semantic embedding of items using a pre-trained Sentence-T5 model~\citep{ni2022sentence} and quantize the resulting item embeddings with an RQ-VAE~\citep{lee2022autoregressive}. The RQ-VAE produces a three-digit hierarchical SID for each item. Besides, we append a fourth anti-collision digit to ensure that every item has a unique SID. 
For the LoopGR backbone, we instantiate the Encoder with T5 Encoder~\citep{raffel2020exploring} and employ a plain Transformer decoder~\citep{vaswani2017attention} as the parameter-shared Decoder.

\begin{table}[t]
\centering
\caption{Performance comparison among baselines and LoopGR}\label{tab:1}
\scriptsize
\setlength{\tabcolsep}{1.5pt}
\begin{tabular}{@{}c|cccc|cccc|cccc@{}}
    \toprule
    \multirow{2}{*}{\textbf{Model}} & \multicolumn{4}{c|}{\textbf{Beauty}} & \multicolumn{4}{c|}{\textbf{Sports and Outdoors}} & \multicolumn{4}{c}{\textbf{CDs and Vinyl}} \\
    \cmidrule(r){2-5} \cmidrule(lr){6-9} \cmidrule(l){10-13}
    & \textbf{R@5} & \textbf{N@5} & \textbf{R@10} & \textbf{N@10} & \textbf{R@5} & \textbf{N@5} & \textbf{R@10} & \textbf{N@10} & \textbf{R@5} & \textbf{N@5} & \textbf{R@10} & \textbf{N@10} \\
    \midrule
    GRU4Rec  & 0.0164 & 0.0099 & 0.0283 & 0.0137 & 0.0129 & 0.0086 & 0.0204 & 0.0110 & 0.0195 & 0.012 & 0.0353 & 0.0171 \\
    BERT4Rec & 0.0203 & 0.0124 & 0.0347 & 0.0170 & 0.0115 & 0.0075 & 0.0191 & 0.0099 & 0.0326 & 0.0201 & 0.0547 & 0.0271 \\
    SASRec   & 0.0387 & 0.0249 & 0.0605 & 0.0318 & 0.0233 & 0.0154 & 0.0350 & 0.0192 & 0.0351 & 0.0177 & 0.0619 & 0.0263 \\
    FDSA     & 0.0267 & 0.0163 & 0.0407 & 0.0208 & 0.0182 & 0.0122 & 0.0288 & 0.0156 & 0.0226 & 0.0137 & 0.0378 & 0.0186 \\
    S3-Rec   & 0.0387 & 0.0244 & 0.0647 & 0.0327 & 0.0251 & 0.0161 & 0.0385 & 0.0204 & 0.0213 & 0.013 & 0.0375 & 0.0182 \\
    VQ-Rec   & 0.0434 & 0.0311 & 0.0741 & 0.0372 & 0.0181 & 0.0132 & 0.0251 & 0.0154 & 0.0314 & 0.0209 & 0.0485 & 0.0264 \\
    \midrule
    TIGER       & 0.0454 & 0.0321 & 0.0648 & 0.0384 & 0.0264 & 0.0181 & 0.0400 & 0.0225 & 0.0492 & 0.0329 & 0.0748 & 0.0411 \\
    HSTU        & 0.0469 & 0.0314 & 0.0704 & 0.0389 & 0.0258 & 0.0165 & 0.0414 & 0.0215 & 0.0417 & 0.0275 & 0.0638 & 0.0346 \\
    ActionPiece & 0.0511 & 0.0340 & 0.0775 & 0.0424 & 0.0316 & 0.0205 & \first{0.0500} & \second{0.0264} & \first{0.0544} & \second{0.0359} & \first{0.0830} & \first{0.0451} \\
    RPG         & 0.0550 & 0.0381 & \second{0.0809} & \second{0.0464} & 0.0314 & \second{0.0216} & 0.0463 & 0.0263 & 0.0498 & 0.0338 & 0.0735 & 0.0415 \\
    \midrule
    LoopGR      & \first{0.0563} & \first{0.0389} & 0.0799 & \first{0.0465} & \first{0.0333} & \first{0.0223} & \second{0.0495} & \first{0.0275} & \second{0.0537} & \first{0.0363} & \second{0.0799} & \second{0.0447} \\
    LoopGR-RTSD & \second{0.0561} & \second{0.0383} & \first{0.0813} & \second{0.0464} & \second{0.0325} & 0.0213 & 0.0484 & \second{0.0264} & 0.0516 & 0.0346 & 0.0789 & 0.0434 \\
    \bottomrule
\end{tabular}
\end{table}

\subsection{Overall Performance}
Table~\ref{tab:1} compares LoopGR with conventional sequential recommenders and recent generative recommendation methods. Overall, generative approaches achieve stronger performance across the three datasets, though several conventional methods remain competitive on individual metrics.
Generative methods exhibit stronger and more consistent performance, demonstrating the benefit of directly generating compositional item identifiers. \textbf{LoopGR and LoopGR (RTSD) achieve the best result on 8 of the 12 evaluation settings.} In particular, LoopGR improves over the strongest baseline by $2.4\%$ in Recall@5 and $2.1\%$ in NDCG@5 for Beauty, and by $5.4\%$ in Recall@5 and $4.2\%$ in NDCG@10 for Sports and Outdoors.

For generative baselines, ActionPiece and RPG employ more elaborate identifier or decoding designs. ActionPiece constructs context-dependent action tokens, whereas RPG adopts long unordered SIDs together with parallel prediction and graph-constrained decoding. 
In contrast, LoopGR adopts a standard SID construction using RQ-VAE, while improving model capacity through a tailored hierarchical recurrent depth for generative recommendation. LoopGR outperforms RPG on 11 of 12 metrics and ActionPiece on 8 of 12 metrics. Moreover, LoopGR (RTSD) remains competitive while using fewer refinement steps and obtains the best Recall@10 on Beauty. It validates the effectiveness of trajectory self-distillation in iterative reasoning for a lower-computation model.

\begin{wraptable}{r}{0.6\textwidth}
\vspace{-13px}
\caption{Offline performance comparison on the industrial dataset.}\label{tab:2}
\scriptsize
\setlength{\tabcolsep}{5pt}
\begin{tabular}{lccccc}
    \toprule
    \textbf{Model} & \textbf{R@5} & \textbf{R@10} & \textbf{R@20} & \textbf{R@50} & \textbf{R@100} \\
    \midrule
    Lazy Decoder & 0.5180 & 0.5796 & 0.6383 & 0.6948 & 0.6990 \\
    TIGER        & 0.5465 & 0.6125 & 0.6668 & 0.7137 & 0.7164 \\
    BARGE        & 0.5477 & 0.6277 & 0.6914 & 0.7469 & 0.7496 \\
    LoopGR       & \first{0.5691} & \first{0.6458} & \first{0.7115} & \first{0.7627} & \textbf{0.7675} \\
    \bottomrule
\end{tabular}
\end{wraptable}
As shown in Table~\ref{tab:2}, LoopGR consistently outperforms the generative recommendation baselines that are used in Tencent's recommendation system. The plain Lazy Decoder-only baseline is highly dependent on a carefully designed context processor, and this architecture may lead to suboptimal performance without an additional post-training phase for preference alignment. BARGE further improves performance through several architectural enhancements, including Item Context-Aware Attention, Hierarchical Path Reranking and Dual-path Decoding. These designs are largely orthogonal to LoopGR and can be incorporated into our framework. LoopGR provides a modular Encoder-Decoder-Updater backbone in which components developed for existing generative recommenders can be integrated in a plug-and-play manner.

\begin{figure*}[t]
    \centering
    \begin{subfigure}[b]{\textwidth}
         \centering
         \includegraphics[width=\textwidth]{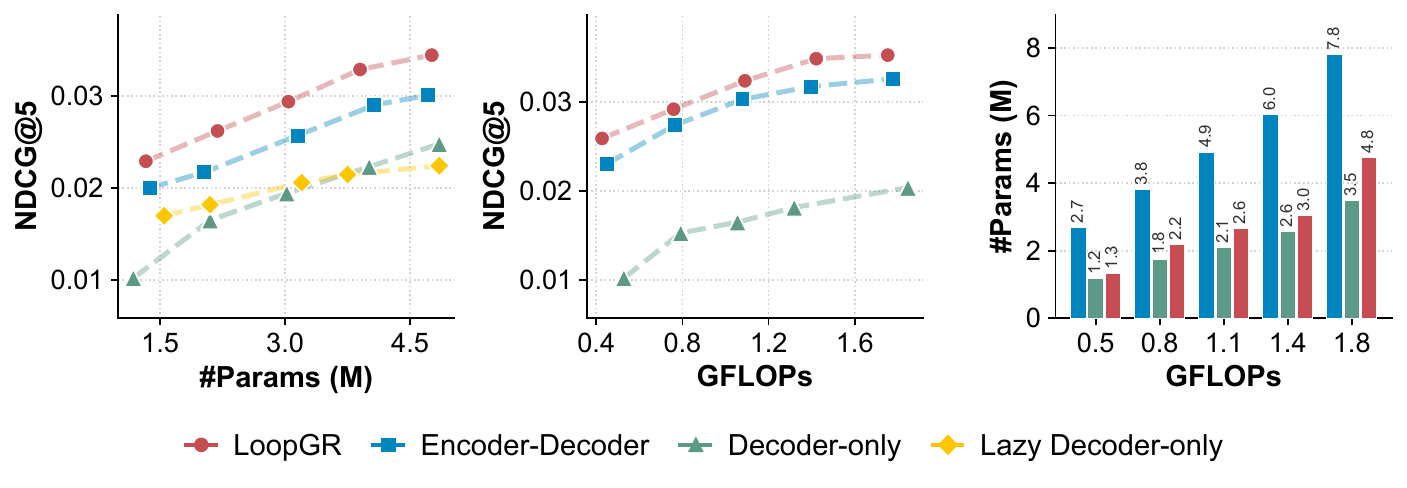}
         \caption{Model scaling under varying parameter and computation budgets on CDs.}
    \end{subfigure}
    \begin{subfigure}[b]{\textwidth}
         \centering
         \includegraphics[width=\textwidth]{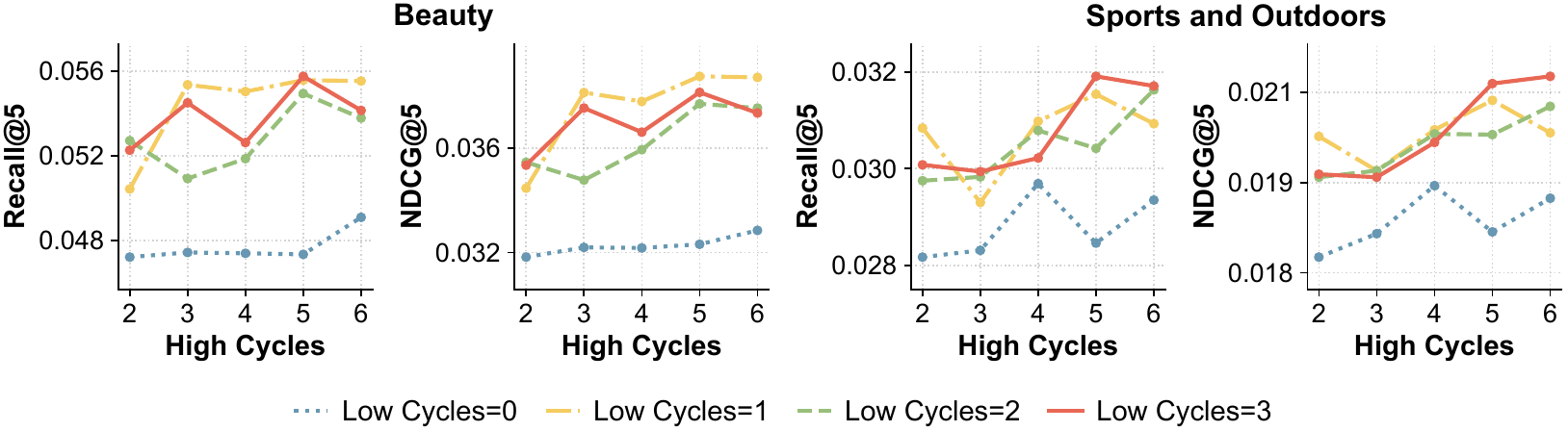}
         \caption{Effect of high-level cycles $T_H$ and low-level cycles $T_L$ on Beauty and Sports.}
    \end{subfigure}
\caption{Architectural comparison across model size, computation and recurrent depth scaling.}
\label{fig:3}
\end{figure*}

\subsection{Architecture Comparison}

To further investigate model size and recurrent-depth scaling, we compare LoopGR with three representative generative backbones, including Encoder-Decoder, Decoder-only and Lazy Decoder-only baselines. For parameter scaling, we fix $T_H=3$ and $T_L=1$ to compare LoopGR with the baselines at comparable parameter counts. For computation scaling, we search over encoder/decoder depths and $T_{H}/T_{L}$ to report the best configuration at comparable GFLOPs. Lazy Decoder-only is excluded from computation scaling because its computation is substantially lower than that of the other architectures. For hierarchical recurrence analysis, we fix both encoder and decoder to one layer and vary $T_H$ and $T_L$. More details about the inference latency are provided in Appendix~\ref{apdx:2}.

\paragraph{Parameter Scaling.}
As shown in Figure~\ref{fig:3} (a), LoopGR achieves higher NDCG@5 than the alternative architectures at similar parameter counts. Its performance continues to improve as the model size increases, while recurrent computation provides an additional dimension for increasing effective depth without introducing new parameters at every step. Consequently, its performance continues to improve beyond Encoder-Decoder as the parameter budget increases. In contrast, Lazy Decoder-only exhibits a flatter scaling curve because the historical context remains fixed and additional parameters are devoted primarily to target decoding.
These results suggest that hierarchical recurrence provides a more parameter-efficient scaling strategy for SID-based generative recommendation.

\paragraph{Computation Scaling.}
LoopGR achieves the highest NDCG@5 across the different computation levels while requiring substantially fewer parameters than Encoder-Decoder. At approximately $0.5$ GFLOPs, LoopGR uses only $1.3$M parameters, less than half of the $2.7$M parameters required by Encoder-Decoder.
Notably, with a parameter budget close to Decoder-only, LoopGR still outperforms the considerably larger Encoder-Decoder model. This demonstrates that additional computation can be introduced through parameter reuse rather than proportional model expansion, yielding a more favorable trade-off among accuracy, computation and model size.

\paragraph{Effect of Hierarchical Recurrent Depth.}
We further vary the high-level depth $T_H$ and low-level depth $T_L$ to examine their respective roles. Figure~\ref{fig:3} (b) shows that removing low-level recurrence $T_L=0$ significantly degrades performance, while introducing even a small number of updater cycles improves both Recall@5 and NDCG@5. It highlights the importance of dynamically refining the historical context between successive decoding cycles. 
Increasing $T_H$ generally brings further gains, although the gains become non-monotonic at larger depths. On Sports, increasing both $T_H$ and $T_L$ yields more pronounced performance gains. This indicates that deeper hierarchical recurrence becomes increasingly beneficial as the recommendation scale grows.
Overall, these results support the hierarchical design of LoopGR, where high-level cycles progressively refine target prediction while low-level cycles adapt the conditioning context between decoding stages.

\subsection{Analysis of Loop Dynamics}
\begin{figure*}[t]
    \centering
    \includegraphics[width=\textwidth]{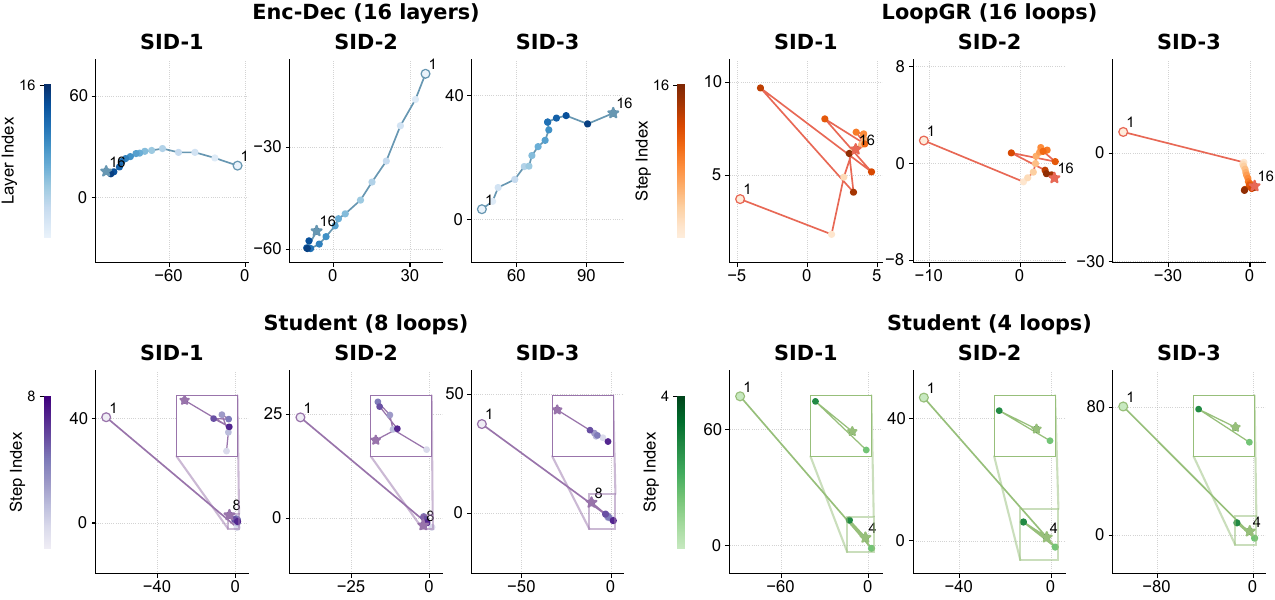}
    \caption{Latent space trajectories for different SIDs.}
\label{fig:4}
\end{figure*}

In this section, we visualize intermediate representation trajectories in latent space across recurrent steps. We focus on two questions: \textbf{(Q1)} how do the latent dynamics of LoopGR differ from those of a conventional Encoder-Decoder? and \textbf{(Q2)} how does trajectory distillation reshape these dynamics under a limited loop budget? 
We collect the intermediate representations at different SID positions from a 16-layer Encoder-Decoder, a 16-loop LoopGR teacher and distilled students with 8 and 4 loops. For each SID position, we project the intermediate states onto the first two principal components to trace their evolution. Figure~\ref{fig:4} provides a visualization of these trajectories. 

\paragraph{Q1: How do the latent dynamics of LoopGR differ from a conventional counterpart?}
As depicted in Figure~\ref{fig:4}, the Encoder-Decoder exhibits relatively smooth layer-wise evolution. Its adjacent representations progressively move along a largely continuous trajectory. In contrast, LoopGR undergoes substantially larger transitions in its early cycles and subsequently enters a more localized refinement regime. 
In the experiments, LoopGR exhibits clear position-dependent dynamics that are characterized by stronger oscillations at earlier SID positions and faster convergence at later SID positions.
These non-monotonic patterns resemble the structured latent trajectories observed in previous Looped Transformers~\citep{geiping2026scaling, wang2025hrm}. Unlike the progressive layer-wise transformations of the Encoder-Decoder, LoopGR repeatedly revises its latent state across cycles, yielding a more dynamic and position-dependent refinement process.

\paragraph{Q2: What does trajectory distillation learn?}
Figure~\ref{fig:4} shows that reducing the loop budget through distillation does not simply truncate the teacher trajectory. The 8-loop student exhibits a much larger transition at the beginning of recurrent computation, followed by progressively smaller updates toward its terminal state. This tendency becomes even more pronounced for the 4-loop student, whose trajectory is dominated by the initial transition.
Here, we propose an intuitive interpretation that the distillation objective encourages a low-budget model to approximate a longer refinement process with fewer recurrent transitions rather than reproducing every intermediate teacher state. Consequently, the teacher and student exhibit complementary computational behaviors. The long-budget teacher develops a richer refinement trajectory, whereas the distilled student learns a more direct trajectory toward the terminal prediction under a limited inference budget.

\subsection{Ablation Study}
\label{sec:ablation_study}
We conduct ablation studies on two key design choices of LoopGR, namely the context updater and the training objective.
For the updater, we replace the default GRU with three sequence modeling methods, including token mixing, linear attention, vanilla attention and an Identity variant that leaves the context representations unchanged. Furthermore, we compare per-step supervision, gated exit, Intra-Loop Self-Distillation \citep[ILSD;][]{goyal2026elt} and Recurrent Trajectory Self-Distillation (RTSD). At inference, the other alternative training objectives use the full budget and RTSD distills LoopGR into a student operating at half of the teacher’s budget.

Table~\ref{tab:3} shows that the Mixing baseline achieves the best Recall@5, Recall@10 and NDCG@10 on Sports. In contrast, replacing the updater with Identity consistently degrades performance, with an average drop of about $8\%$ relative to the learnable updater variants. This confirms that the gain primarily comes from dynamically updating the conditioning context rather than from a particular sequence operator, making the updater design largely architecture-agnostic. 
Averaged over the eight reported metrics, per-step supervision, gated exit and ILSD fall $9.7\%$, $13.2\%$ and $9.4\%$ below the full-budget LoopGR, respectively. RTSD reduces this gap to only $3.2\%$. We suggest that direct supervision at intermediate cycles can induce representation stagnation. It causes subsequent updates to diminish before the full recurrent budget is exploited. RTSD instead separates trajectory learning from compression and better preserves performance under a reduced budget.

\begin{table}[t]
\centering
\scriptsize
\caption{Ablation study of different updater architectures and training objectives. The best and second-best results in each column are highlighted in bold and underlined, respectively.}\label{tab:3}
\begin{tabular}{@{}l|cccc|cccc@{}}
    \toprule
    \multirow{2}{*}{\textbf{Model}} & \multicolumn{4}{c|}{\textbf{Sports and Outdoors}} & \multicolumn{4}{c}{\textbf{CDs and Vinyl}} \\
    \cmidrule(r){2-5} \cmidrule(lr){6-9}
    & \textbf{R@5} & \textbf{N@5} & \textbf{R@10} & \textbf{N@10} & \textbf{R@5} & \textbf{N@5} & \textbf{R@10} & \textbf{N@10} \\
    \midrule
    LoopGR & 0.0333 & \first{0.0223} & 0.0495 & 0.0275 & \first{0.0537} & \first{0.0363} & \first{0.0799} & \first{0.0447} \\
    \midrule
    w. Identity     & 0.0311 & 0.0197 & 0.0489 & 0.0254 & 0.0502 & 0.0324 & 0.0730 & 0.0400 \\
    w. Mixing       & \first{0.0342} & \second{0.0222} & \first{0.0531} & \first{0.0283} & 0.0527 & 0.0352 & \second{0.0792} & 0.0436 \\
    w. Linear Attn  & \second{0.0338} & 0.0219 & 0.0516 & 0.0276 & \first{0.0537} & 0.0354 & 0.0789 & 0.0435 \\
    w. Vanilla Attn & 0.0333 & 0.0217 & \second{0.0520} & \second{0.0277} & \second{0.0533} & \second{0.0362} & 0.0768 & \second{0.0437} \\
    \midrule
    w. Per-Step Supervision & 0.0280 & 0.0186 & 0.0424 & 0.0232 & 0.0516 & 0.0348 & 0.0771 & 0.0430 \\
    w. Gated Exit           & 0.0264 & 0.0174 & 0.0406 & 0.0220 & 0.0502 & 0.0338 & 0.0754 & 0.0420 \\
    w. ILSD                 & 0.0290 & 0.0192 & 0.0427 & 0.0236 & 0.0512 & 0.0347 & 0.0752 & 0.0424 \\
    w. RTSD                 & 0.0325 & 0.0213 & 0.0484 & 0.0264 & 0.0516 & 0.0346 & 0.0789 & 0.0434 \\
    \bottomrule
\end{tabular}
\end{table}

\begin{table}[htbp]
\centering
\caption{Offline performance on the industrial dataset with different context updater designs.}
\label{tab:4}
\small
\setlength{\tabcolsep}{5pt}
\begin{tabular}{lccccc}
    \toprule
    \textbf{Model} & \textbf{R@5} & \textbf{R@10} & \textbf{R@20} & \textbf{R@50} & \textbf{R@100} \\
    \midrule
    Lazy Decoder & 0.5180 & 0.5796 & 0.6383 & 0.6948 & 0.6990 \\
    TIGER        & 0.5465 & 0.6125 & 0.6668 & 0.7137 & 0.7164 \\
    BARGE        & 0.5477 & 0.6277 & 0.6914 & 0.7469 & 0.7496 \\
    \midrule
    LoopGR (w. Identity) & 0.5513 & 0.6188 & 0.6789 & 0.7334 & 0.7355 \\
    LoopGR (w. GRU) & \second{0.5626} & \second{0.6383} & \second{0.7007} & \second{0.7563} & \second{0.7585} \\
    LoopGR (w. Linear Attn) & 0.5556 & 0.6245 & 0.6854 & 0.7395 & 0.7422 \\
    LoopGR (w. Mixing) & \first{0.5691} & \first{0.6458} & \first{0.7115} & \first{0.7627} & \first{0.7675} \\
    \bottomrule
\end{tabular}
\end{table}

We further evaluate different LoopGR variants on the industrial dataset by varying the design of the context updater. As shown in Table~\ref{tab:4}, LoopGR achieves the best performance across all evaluation metrics, with maximum relative improvements of 11.47\%, 7.13\% and 3.91\% over Lazy Decoder, TIGER and BARGE, respectively. Notably, its gain over TIGER becomes increasingly pronounced at larger retrieval cutoffs, suggesting that LoopGR is particularly effective at improving recall over a broader candidate set. These results further demonstrate that LoopGR can effectively incorporate different lightweight context-updating operators while retaining strong performance in large-scale industrial recommendation.

\section{Conclusion}

In this paper, we introduce \textbf{LoopGR}, a parameter-efficient framework for generative recommendation that decouples computational depth from parameter growth through hierarchical recurrent computation. 
By recurrently reusing a decoder and a context updater, LoopGR increases effective depth while dynamically refining the historical context during SID generation. We further develop recurrent trajectory self-distillation to better balance recommendation performance and inference efficiency under hard recurrent budgets.
Extensive experiments show that LoopGR consistently outperforms strong baselines and achieves better performance than representative Encoder-Decoder architectures with up to $52\%$ fewer parameters. These results demonstrate the potential of recurrent depth as a complementary scaling dimension beyond model size for generative recommendation.

\bibliography{reference}
\bibliographystyle{conference_bib}


\appendix
\section{Appendix}

\subsection{Experimental Details}\label{apdx:1}
We conduct experiments on three public Amazon review datasets, including Beauty, Sports and Outdoors and CDs and Vinyl. As summarized in Table~\ref{tab:dataset_statistics}, the datasets span different scales and interaction densities. Beauty contains 22363 users, 12101 items and 198502 interactions, while Sports and Outdoors includes 35598 users, 18357 items and 296337 interactions. CDs and Vinyl is the largest dataset with 75258 users, 64443 items and over one million interactions. The average interaction lengths are 8.88, 8.32 and 14.58 for Beauty, Sports and CDs, respectively.

\begin{table}[htbp]
\centering
\caption{Statistics of the datasets. The average length denotes the average number of interactions per user.}
\label{tab:dataset_statistics}
\setlength{\tabcolsep}{8pt}
\begin{tabular}{lcccc}
\toprule
\textbf{Dataset} &
\textbf{\#Users} &
\textbf{\#Items} &
\textbf{\#Interactions} &
\textbf{Avg. Length} \\
\midrule
Beauty & 22,363 & 12,101 & 198,502   & 8.88  \\
Sports & 35,598 & 18,357 & 296,337   & 8.32  \\
CDs    & 75,258 & 64,443 & 1,097,592 & 14.58 \\
\bottomrule
\end{tabular}
\end{table}

\subsection{Efficiency under Compressed Recurrent Trajectories}\label{apdx:2}
We further evaluate whether RTSD can preserve recommendation quality when the recurrent trajectory is compressed to fewer cycles. Reducing the teacher from 16 cycles to 8 and 4 cycles decreases computation from 17.408 GFLOPs to 8.415 and 3.918 GFLOPs, respectively. The corresponding inference latency drops from 211.08 ms to 104.01 ms and 49.82 ms under batch size 1, providing approximately 2.0$\times$ and 4.2$\times$ speedups. Despite the substantially shorter trajectories, both students retain comparable recommendation accuracy. Notably, the 4-cycle student surpasses the 16-cycle teacher on all four metrics for Sports and remains competitive on Beauty, while the 8-cycle student achieves the best Recall@10 and NDCG@10 on Beauty. These results show that RTSD can compress a long recurrent trajectory into substantially fewer cycles without proportionally sacrificing predictive quality, enabling a more favorable trade-off between recurrent depth and inference efficiency.

\begin{table*}[htbp]
\centering
\caption{
Performance-efficiency comparison between the full-budget teacher and students with reduced recurrent cycles. Inference latency is measured with batch size 1.
}
\label{tab:rtsd_efficiency}
\scriptsize
\setlength{\tabcolsep}{1.5pt}
\begin{tabular}{c|cc|cccc|cccc}
    \toprule
    \multirow{2}{*}{\textbf{Model}} & \multirow{2}{*}{\textbf{GFLOPs}} & \multirow{2}{*}{\textbf{\makecell{Latency \\ (ms)}}} & \multicolumn{4}{c|}{\textbf{Sports and Outdoors}} & \multicolumn{4}{c}{\textbf{Beauty}} \\
    \cmidrule(lr){4-7} \cmidrule(lr){8-11}
    & & & \textbf{R@5} & \textbf{N@5} & \textbf{R@10} & \textbf{N@10} & \textbf{R@5} & \textbf{N@5} & \textbf{R@10} & \textbf{N@10} \\
    \midrule
    LoopGR (Teacher, 16-loop) & 17.408 & 211.08 & 0.0304 & 0.0199 & 0.0456 & 0.0248 & 0.0531 & \textbf{0.0369} & 0.0775 & 0.0448 \\
    LoopGR-RTSD (8-loop) & 8.415 & 104.01 & 0.0295 & 0.0195 & 0.0465 & 0.0250 & 0.0530 & \textbf{0.0369} & \textbf{0.0791} & \textbf{0.0453} \\
    LoopGR-RTSD (4-loop) & \textbf{3.918} & \textbf{49.82} & \textbf{0.0315} & \textbf{0.0209} & \textbf{0.0479} & \textbf{0.0261} & \textbf{0.0534} & 0.0368 & 0.0789 & 0.0450 \\
    \bottomrule
\end{tabular}
\end{table*}

\subsection{Additional Visualization of Recurrent Trajectories}\label{apdx:3}
\begin{figure*}[htbp]
    \centering
    \begin{subfigure}[t]{\textwidth}
        \centering
        \includegraphics[width=0.9\linewidth]{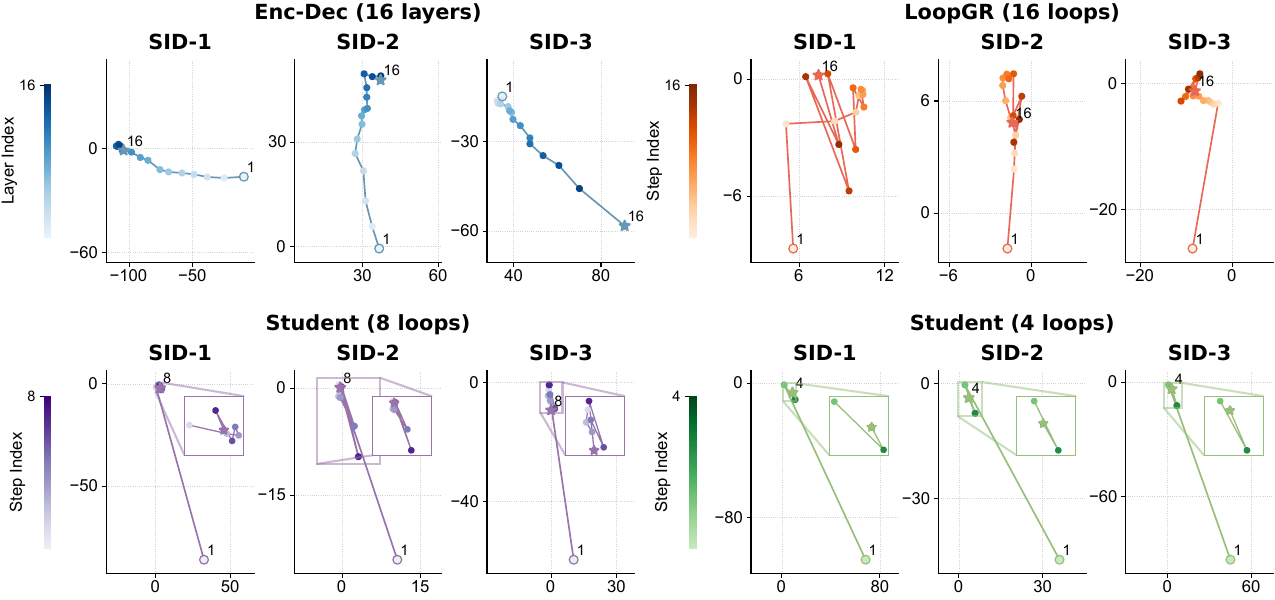}
        \caption{Sample 0}
        \label{fig:traj_sample0}
    \end{subfigure}
    \hfill
    \begin{subfigure}[t]{\textwidth}
        \centering
        \includegraphics[width=0.9\linewidth]{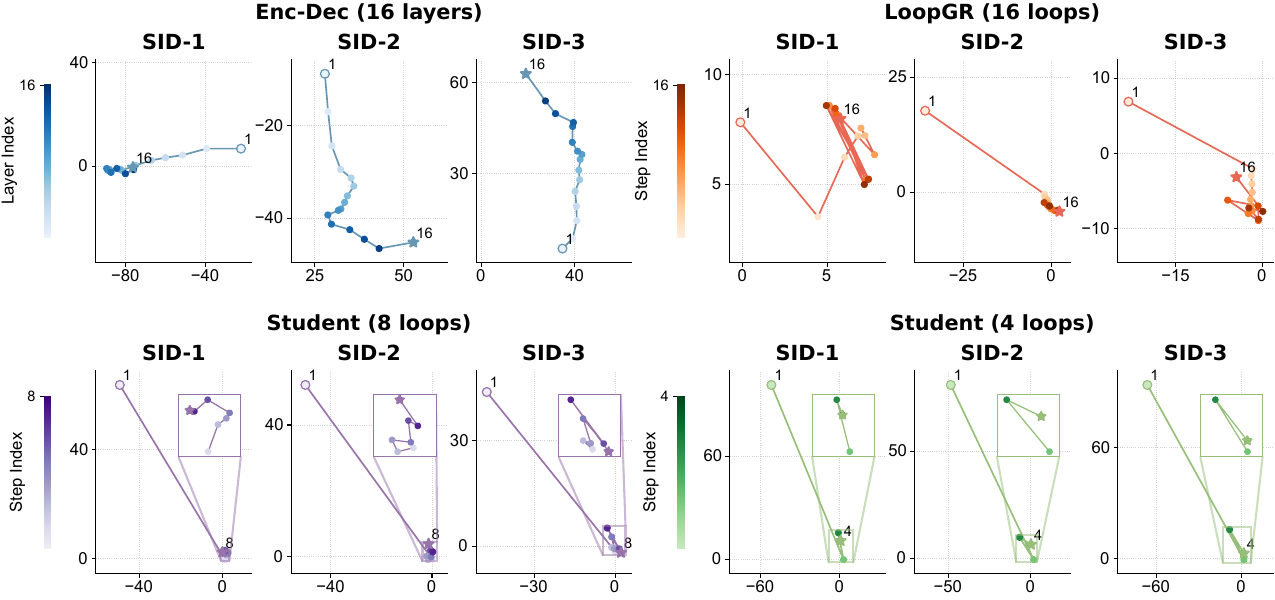}
        \caption{Sample 1}
        \label{fig:traj_sample1}
    \end{subfigure}
    \hfill
    \begin{subfigure}[t]{\textwidth}
        \centering
        \includegraphics[width=0.9\linewidth]{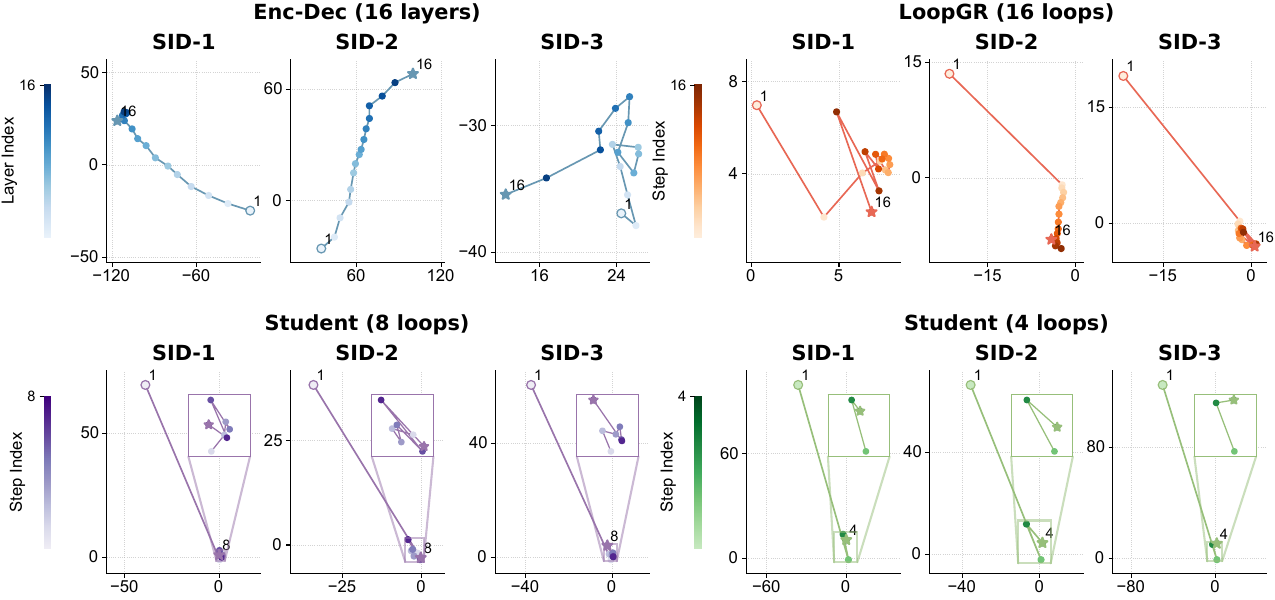}
        \caption{Sample 2}
        \label{fig:traj_sample2}
    \end{subfigure}

    \caption{Visualization of recurrent trajectories for different samples.}
    \label{fig:trajectory_samples}
\end{figure*}

We provide additional trajectory visualizations to complement the analysis in the main text. Figure~\ref{fig:trajectory_samples} shows representative examples across the three SID positions, comparing the 16-layer Encoder-Decoder, the 16-loop LoopGR teacher, and the distilled students with 8 and 4 recurrent cycles. Across different samples, LoopGR exhibits iterative refinement throughout recurrent computation, while the distilled students reach similar terminal regions through substantially shorter trajectories. These examples show that the distilled students reach their terminal states through shorter trajectories.

\subsection{Implementation Details of the Context Updater}\label{apdx:4}

As discussed in the main text, the context updater $\operatorname{UPD}$ is model-agnostic, allowing different architectures to be incorporated into the same recurrent refinement framework. Here, we categorize the updaters into token-level and item-level variants according to their modeling granularity. Token-level updaters operate directly on individual SID token representations. Item-level updaters group the $H$ SID tokens of each historical item and refine the resulting item representations. Each historical item is represented by $H$ SID tokens. We index the context representations by item and hierarchy, and the representation of the $h$-th SID token of the $j$-th historical item is denoted as $\mathbf{m}_{j,h}\in\mathbb{R}^{d}$, where $j=1,\dots,n$ and $h=1,\dots,H$. For brevity, we omit the cycle indices.

\paragraph{Token-level Updaters.}
Token-level updaters refine the context directly at the granularity of SID tokens. As a representative example, linear attention~\citep{katharopoulos2020transformers} enables each token to aggregate information from the entire history through a kernel-based attention mechanism. The prediction-aware query $\mathbf{q}$ is injected only into the attention query branch, while keys and values are derived from the conditioned context. For a single attention head, the query, key and value projections for the $h$-th SID token of historical item $i_j$ are
\begin{equation}
    \mathbf{Q}_{j,h} = \mathbf{W}_Q\big(\widehat{\mathbf{m}}_{j,h}+\mathbf{q}\big),\qquad
    \mathbf{K}_{j,h} = \mathbf{W}_K\widehat{\mathbf{m}}_{j,h},\qquad
    \mathbf{V}_{j,h} = \mathbf{W}_V\widehat{\mathbf{m}}_{j,h},
\end{equation}
where $\widehat{\mathbf{m}}_{j,h}$ denotes the conditioned representation of the $h$-th SID token of item $j$. $\mathbf{W}_Q$, $\mathbf{W}_K$ and $\mathbf{W}_V$ are learnable projections. Using the positive feature map $\phi(\cdot)=\operatorname{ELU}(\cdot)+1$, the attention output can be computed by exploiting the associative property of matrix multiplication,
\begin{equation}
    \mathbf{o}_{j,h} =
    \frac{\sum_{a,b}\big[\phi(\mathbf{Q}_{j,h})^{\top}\phi(\mathbf{K}_{a,b})\big]\mathbf{V}_{a,b}}
    {\sum_{a,b}\phi(\mathbf{Q}_{j,h})^{\top}\phi(\mathbf{K}_{a,b})}
    =
    \frac{\big[\sum_{a,b}\mathbf{V}_{a,b}\phi(\mathbf{K}_{a,b})^{\top}\big]\phi(\mathbf{Q}_{j,h})}
    {\phi(\mathbf{Q}_{j,h})^{\top}\sum_{a,b}\phi(\mathbf{K}_{a,b})},
\end{equation}
Linear attention effectively avoids explicit pairwise attention scores with a computational complexity of $\mathcal{O}(L_u d^2)$.
Token-level updaters can also be instantiated as an MLP or an identity function, applied independently to each query-conditioned SID token representation. The MLP transforms each token through a feed-forward network, whereas the identity function leaves its input unchanged.

\paragraph{Item-level Updaters.}
Item-level updaters treat the $H$ SID tokens of each item as a group rather than independent modeling units. They first organize token representations into item-level inputs. After refinement, they map the resulting updates back to the SID sequence. Specifically, the GRU updater uses mean pooling to summarize each item's SID tokens as a vector. These vectors are projected and combined with the prediction-aware query before a bidirectional GRU~\citep{chung2014empirical}, which processes the history in chronological and reverse chronological order. The forward and backward states are concatenated and projected to produce an item-level update $\mathbf{u}_j\in\mathbb{R}^{d}$, which is then copied to all $H$ SID tokens of that item. The pooling and broadcast operations are
\begin{align}
    \text{Pooling:}\quad \bar{\mathbf{m}}_j &= \frac{1}{H}\sum_{h=1}^{H}\widehat{\mathbf{m}}_{j,h}, \\
    \text{Broadcast:}\quad \widetilde{\mathbf{m}}_{j,h} &= \mathbf{u}_j,
\end{align}
With the GRU hidden dimension per direction set to $d$, the sequence modeling and projection cost is $\mathcal{O}(n d^2)$.

The Mixing updater preserves the hierarchy-specific representations by grouping each item's SID tokens into an $H\times d$ matrix. After query injection, token mixing rearranges channel groups across the $H$ tokens, so each output slot receives features from every SID token of the same item~\citep{zhu2025rankmixer}. Hierarchy-specific feed-forward networks refine these mixed features, followed by an output projection to produce $\mathbf{U}_j\in\mathbb{R}^{H\times d}$. Unlike the shared update in the GRU updater, each row of this matrix is mapped back to its corresponding SID token:
\begin{align}
    \text{Grouping:}\quad \widehat{\mathbf{M}}_j &= \big[\widehat{\mathbf{m}}_{j,1}^{\top};\dots;\widehat{\mathbf{m}}_{j,H}^{\top}\big], \\
    \text{Broadcast:}\quad \widetilde{\mathbf{m}}_{j,h} &= [\mathbf{U}_j]_{h,:}^{\top},
\end{align}
For a fixed number of mixing layers, the computational complexity is $\mathcal{O}(L_u d^2)$.

The two variants provide complementary inductive biases. The GRU updater captures bidirectional dependencies across items but compresses their internal SID structure through mean pooling, whereas the Mixing updater preserves and refines hierarchy-specific information within each item without introducing additional inter-item communication.

\end{document}